\documentclass[conference]{IEEEtran}
\IEEEoverridecommandlockouts
\usepackage{cite}
\usepackage{amssymb,amsfonts}
\usepackage{algorithmic}
\usepackage[ruled]{algorithm2e}
\usepackage{textcomp}
\usepackage{xcolor}
\usepackage{amsmath,graphicx}
\usepackage{booktabs}
\usepackage{subfigure}

\def\BibTeX{{\rm B\kern-.05em{\sc i\kern-.025em b}\kern-.08em
    T\kern-.1667em\lower.7ex\hbox{E}\kern-.125emX}}
\begin{document}

\title{MetaSpeech: Speech Effects Switch Along with Environment for Metaverse
}

% \and
% \and
% \and
% \and
% \and

\author{\IEEEauthorblockN{Xulong Zhang, Jianzong Wang$^\ast$\thanks{$^\ast$Corresponding author: Jianzong Wang, jzwang@188.com.}, Ning Cheng, Jing Xiao}
\IEEEauthorblockA{\textit{Ping An Technology (Shenzhen) Co., Ltd.}}
}

\maketitle

\begin{abstract}
    Metaverse expands the physical world to a new dimension, and the physical environment and Metaverse environment can be directly connected and entered. Voice is an indispensable communication medium in the real world and Metaverse. Fusion of the voice with environment effects is important for user immersion in Metaverse. In this paper, we proposed using the voice conversion based method for the conversion of target environment effect speech. The proposed method was named MetaSpeech, which introduces an environment effect module containing an effect extractor to extract the environment information and an effect encoder to encode the environment effect condition, in which gradient reversal layer was used for adversarial training to keep the speech content and speaker information while disentangling the environmental effects. From the experiment results on the public dataset of LJSpeech with four environment effects, the proposed model could complete the specific environment effect conversion and outperforms the baseline methods from the voice conversion task.
\end{abstract}


\section{Introduction}

Metaverse~\cite{lee2021all,ning2021survey,lim2022realizing} is the expansion of the real world in the virtual world. It is digitalization and virtualization based on the physical world so that people can carry out their daily work and entertainment in a more convenient way. People can enter Metaverse at any time from different locations, and instantly enter the same meeting room to start a work discussion meeting on time. Eliminates the physical constraints of the physical world and the constraints of resource-constrained conference rooms~\cite{sparkes2021metaverse}. We can also carry out unrealistic activities in the real world in Metaverse. For example, in Metaverse, it is possible to achieve instantaneous movement free from geographical constraints and production and creation free from material constraints.

Although Metaverse can break many constraints in the real world, Metaverse must also provide a more realistic experience and immersion~\cite{duan2021metaverse}. People can interact and walk around in different environments of Metaverse, such as ending a work meeting from a conference room in Metaverse, entering a gallery in Metaverse to enjoy art parting, or entering a concert hall to enjoy a large-scale symphony. Although the transition of different scenes can be ignored as instant transfer, the immersion needs to allow users to experience different environmental atmospheres in different environments, including visual space and sound.

Voice effects~\cite{ratnarajah2021fast,SuwanaposeeGC22,RamirezWSB21} in different environments will bring different perception effects to listeners according to the size of the space and the material of the environment. Two methods are usually used to add environmental sound effects to the recorded vocal sound in a virtual scene~\cite{martinez2021deep,abs-2006-05584}, one is parameter calculation, and the other is manual tuning. The parameter calculation is to calculate the reflection structure of the sound according to the material and distance in the physical environment to convolve the comb-like shape filter with the vocal sound. Manual tuning~\cite{bridges2020effects} is to adjust or increase certain audio components in the audio based on experience. For different scenes in Metaverse, different sound effects need to be designed to enhance the difference of playback audio, such as enhancing vocals, bass compensation, expanding surround and creating artificial reverberation, \textit{etc.}, to achieve surround feeling, vocals, presence, and other auditory effects are enhanced. 

 An audio effect based on modulation or varying with time relates to an audio processor. The use of delay lines and digital filters can implement many audio effects~\cite{canfield2018group}. Recently the deep model-based method has achieved outperformance on many tasks~\cite{gao2021vocal,zhang2020research}, it also been applied for the generation of audio effects.  Convolutional neural networks and recurrent neural networks are combined to model audio special effects. While the handcraft audio effects or the learned audio effects can be directly applied to the vocal to achieve the specific environmental effects. But there needs the clean vocal of the speech such as the studio or soundproof room. This is not easy for the user to get access to Metaverse anywhere they want.

In this paper, we proposed the framework of effect conversion to remove the source effect and replace it with the specific target effect. For the effect conversion, we disentangle the speech and environment with two separate representations. With the reference speech in the target environment, we extract the environment latent and fusion with the speech of the source to decode the generated speech with only the target environment effect. To enhance the naturalness of the generated speech, a variance adaptor was added to the latent representation.   

Our contribution can be concluded as: 1) For Metaverse, we proposed the speech effects switch method by the framework of effect conversion. 2) Disentanglement of the environment effect was modeled as a latent representation of an effect extractor. 3) Variance adaptor was proposed to enhance the naturalness of the generated speech.

% \begin{enumerate}
% \item For Metaverse, we proposed the speech effects switch method by the framework of effect conversion.

% \item Disentanglement of the environment effect was modeled as a latent representation of an effect extractor.

% \item Variance adaptor was proposed to enhance the naturalness of the generated speech.
% \end{enumerate}

\section {Related Works}

% The effect conversion is related with the task of effect generator. As shown in Figure \ref{framework compare}, the left \ref{framework compare a} is a traditional audio effect generator, the right \ref{framework compare b} is the proposed framework of effect conversion.
% \begin{figure}[htbp]
%   \centering
%   \subfigure[effect generator]{
%   \includegraphics[width=0.45\linewidth]{figs/fig-framework-raw.png}
%   \label{framework compare a}
%   }
%   \subfigure[effect conversion]{
%   \includegraphics[width=0.45\linewidth]{figs/fig-framework-our.png}
%   \label{framework compare b}
%   }
%   \caption{The framework of environment effect switch: (a) effect generator and (b) effect conversion}
%     \label{framework compare}
%   \end{figure}

The effect conversion task is similar to the voice conversion (VC)~\cite{tang2022avqvc,ChenSH21,wang2022drvc,ChenWWL21,asru2021tang,HayashiHKT21,asru2021zhang} task in terms of spectrum conversion, both need to do a conversion of the target speech. But in the voice conversion, there is only a need to keep the content the same. The conversion of the environment effect needs to keep the content and the same timbre of the source speaker. To some extent, it can be treated as the same when content not just the information of speech text.

The voice conversion models can be categorized into three main classes, they are GAN based model~\cite{kaneko2017parallel,kameoka2018stargan,kaneko2021maskcyclegan,zhang2021singer}, VAE based model~\cite{hsu2016voice,kameoka2018acvae,zhang2020singing} and encoder-decoder based model~\cite{qian2019autovc,qian2020unsupervised,qian2021global}. Kaneko \textit{et al.}~\cite{kaneko2017parallel} applied the CycleGAN on the voice conversion task, with the cycle consistent loss resolving the need of parallel dataset. This method shows a performance comparably to a parallel VC method. However, the generated speech still has a large gap with the real speech. An enhanced version CycleGAN-VC2 is updated by Kaneko \textit{et al.}~\cite{kaneko2019cyclegan}, which incorporates three new enhancements for the generator, discriminator, and objective separately. However, the two methods are both used for mel-cepstrum conversion, which cannot be directly used for mel-spectrum or spectrum, which has no ability for the modeling of the aperiodicities information. 

While the GAN based VC methods are tough in training and have the poor ability of generalization to the out-of-set speaker. On the other hand, VAE is easier to train. The VAE based model~\cite{hsu2016voice,zhao2022nnspeech} also can commit a conversion on the non-parallel corpora. Through a condition of speaker embedding as an additional input for CVAE~\cite{kameoka2018acvae} can achieve specific conversion of the target speaker. However, CVAE alone often suffers from over-smoothing of the output and cannot guarantee the distribution matching. Qian \textit{et al.}~\cite{qian2019autovc} propose to use an autoencoder with a well-designed bottleneck for the disentanglement of content and speaker style. With only a self-reconstruct loss it can achieve distribution matching style transfer and could perform the zero-shot voice conversion. 

To the environment effect switch task, many environments could be created and the entrance environment to Metaverse could be various. Motivated by the voice conversion methods, we proposed to do a disentanglement of the effect and commit any to any conversion of environment effect.    

\section{Method}
In this section, we first give the overview of the proposed method and then show the detail of the main components. We also introduce the training and inference process of the proposed method for speech environment effect conversion.  

\subsection{Model Overview}
\begin{figure}[htbp]
\centerline{\includegraphics[width=\linewidth]{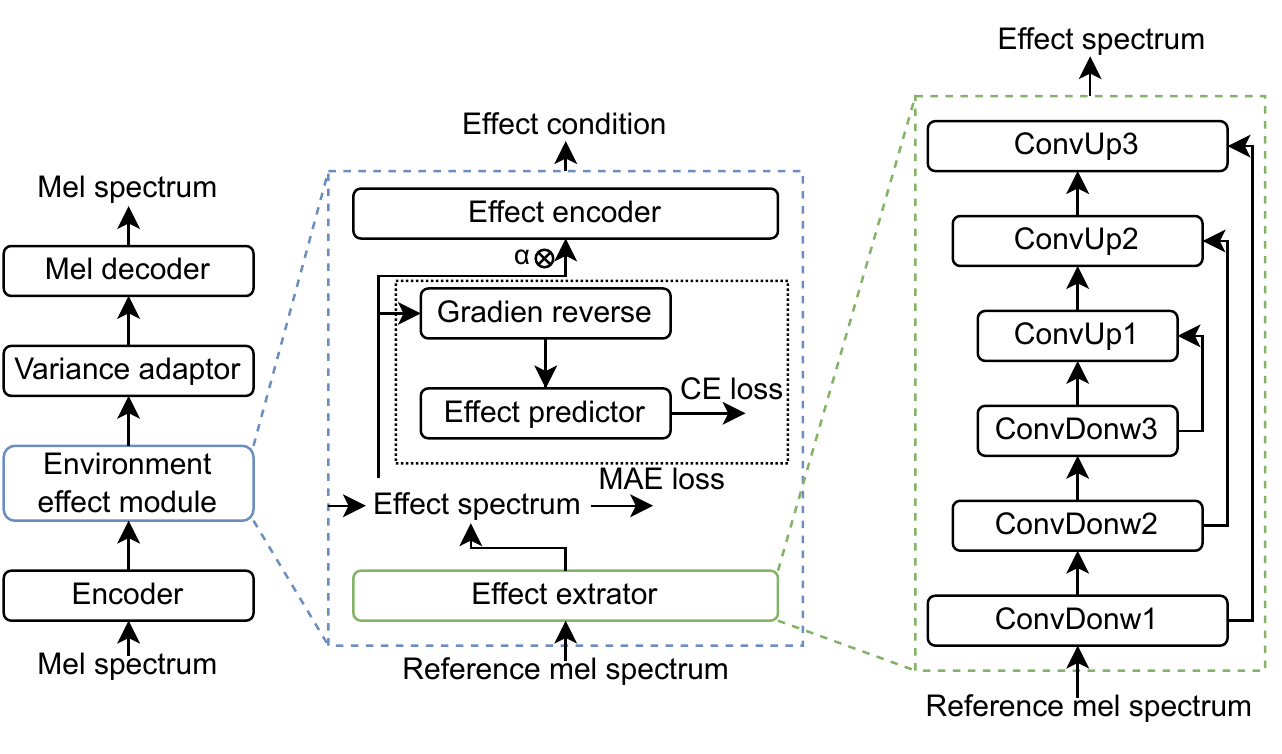}}
\caption{The framework of environment effect conversion}
\label{fig-overview}
\end{figure}
As shown in Figure~\ref{fig-overview}, the main modules include a mel encoder, an environment effect module. The mel encoder is built up with convolutional layers with 1-dimensional convolution along the time axis. The variance adaptor based on the work in~\cite{ren2020fastspeech}, it contains pitch predictor and energy predictor to predict the pitch and energy for naturalness enhancement of the generated target speech. The mel decoder is used for the generation of mel spectrum from the latent variable. A feed-forward transformer is used for the mel decoder. The environment effect module is the core of the speech environment effect conversion, it contains an effect extractor, an effect encoder, and an effect predictor to enhance the extracted effect in an adversarial way. The Environment effect module will be described in detail at \ref{subsec:environment effect module}.  
\subsection{Environment Effect Module}
\label{subsec:environment effect module}
In this section, we will introduce the environment effect module. Trough an effect extractor and an effect predictor with a target-specific gradient reversal layer to enhance the representation of the effect spectrum. Finally, the controllable effect spectrum is embedded as an effect condition to add the speech content for the environment effect conversion. 

\subsubsection{Effect Extractor}
The effect extractor aims to disentangle the effect spectrum $y^{'}$ from the reference mel spectrum $y$. As shown in the right of Figure~\ref{fig-overview}, we proposed to use the architecture of Unet for effect spectrum extractor. There are three convolutional down layers and three convolutional up layers, both up and down layers are used 1-dimensional convolution, and each convolutional layer follows with a batch normalization layer and with the activation of ReLU. The training of the Unet is jointly with the effect spectrum predictor, the classifier can help the end-to-end gradient propagation without a specific label of the target effect spectrum $y^{'}$. 
\subsubsection{Effect Encoder}
The effect encoder combines the source speech content and the extracted reference effect spectrum with a controllable factor $\alpha$ to generate an effect condition of the speech of the target environment.  The effect encoder is built up with a convolution layer, padding and dilation are both 1. For constraints of the same length, during the training, we used two paired data that one is the source and reference spectrum is the same and the other is with the same environment and a fixed max length. 

\subsubsection{Adversarial Classifier}
We set two adversarial classifiers in the environment effect module. One is for the mel encoder output with gradient reversal layer to make the mel encoder without the representation of environment effect. The other one is for the effect extractor to clearly represent the specific effect with gradient reversal on non-target effect samples. In this way, we force the effect encoder to learn the representation related to the specific effect without containing speech content information. Let $x$ be the input source mel spectrum, there is a reconstruct loss for the speech as shown in Equation~\ref{eq:recon loss}.

\begin{equation}
    \label{eq:recon loss}
    L_{recon}=L_{MSE}(x,Dec(Enc(x)+EE(y)+Var(Enc(x))))
\end{equation}
where $L_{MSE}(\cdot,\cdot)$ is the calculation of mean squared error loss, $Dec(\cdot)$ is the mel decoder, $Enc(\cdot)$ is the mel encoder, $EE(\cdot)$ is the effect extractor, $Var(\cdot)$ is the variance adaptor. There are two losses in the variance adaptor as shown in Equation~\ref{eq:loss pitch} and Equation~\ref{eq:loss energy} for the pitch predictor ($L_{pitch}$) and energy predictor ($L_{energy}$) separately.

\begin{equation}
    \label{eq:loss pitch}
    L_{pitch}=L_{MSE}(x_p,PP(Enc(x)))
\end{equation}
\begin{equation}
    \label{eq:loss energy}
    L_{energy}=L_{MSE}(x_e,EP(Enc(x)))
\end{equation}
where $x_p$ and $x_e$ are the pitch and energy of speech $x$ separately. The $PP(\cdot)$ is the pitch predictor, and $EP(\cdot)$ is the energy predictor. There are two adversarial losses $L_{advC}$ and $L_{advE}$ as shown in Equation \ref{eq:loss advc} and Equation \ref{eq:loss adve} for the two classifiers of encoder content and environment effects separately. 

\begin{equation}
    \label{eq:loss advc}
    L_{advC}=L_{ce}(GRL(Enc(x)),x_{ef})
\end{equation}
where $GRL(\cdot)$ is the gredient reversal layer, $x_{ef}$ is the environment class of the speech $x$.
\begin{equation}
    \label{eq:loss adve}
    L_{advE}=L_{ce}(GRL_{non}(EE(y)),y_{ef})
\end{equation}
where $GRL_{non}(\cdot)$ is the gradient reversal layer only work on non target classes, $y_{ef}$ is the environment effect of reference speech. Finally, the total loss $L_{total}$ is the sum of the five losses as shown in Equation \ref{eq:loss total}.

\begin{equation}
    \label{eq:loss total}
    L_{total}=L_{recon}+L_{pitch}+L_{energy}+L_{advC}+L_{advE}
\end{equation}

\subsection{Training and Inference}

 The detailed procedure is shown in Algorithm \ref{al:training and inference}. There are three steps in the model training phase. The same environment effect audios will be used for the training of environment effect extractor by a self reconstruction task. In the second step, the environment classifier was trained by adding a gradient reversal layer, it does reverse the gradient for the specific target of the output of the environment effect extractor. In the third step, the variance adaptor, classifiers, and the encoder-decoder for mel spectrum are jointly trained. During the inference phase, there are mainly three steps for the target environment audio generation. In the first step, the target effect condition was extracted from the reference mel spectrum. In the second step, encoded source audio and the effect condition for the latent vector of the target audio. In the third step, with the latent vector of the target audio to decode. Finally, we use a vocoder of HiFi-GAN to get the audio waveform. 

\begin{algorithm}
\label{al:training and inference}
  \SetAlgoLined
    \textbf{Training phase:}\\
    \KwIn{source mels $x$, reference mels $y$, source pitch $x_p$, source energy $x_e$, source environment $x_{ef}$}
    \KwResult{trained model $f(\cdot)$}
    \textbf{Step 1:} Self reconstruct using the same environment effect reference audio for the effect extractor.\\  
    \textbf{Step 2:} Add gradient reversal layer to train the environment classifier for the output of mel encoder. Specific target to add gradient reversal layer for the environment classifier for the output of effect extractor.\\
    \textbf{Step 3:} Jointly train the variance adaptor, environment classifiers and the encoder decoder for mel spectrum.\\
  \hrulefill\\
  \textbf{Inference phase:}\\
\KwIn{source mel, reference mel}
\KwResult{target environment audio}
\textbf{Step 1:}
Extract environment effect condition from reference mel.\\
\textbf{Step 2:}
Encode the source target and sum it with the effect condition with a controllable weight.\\
\textbf{Step 3:}
Decode the latent variable to mel spectrum. With the vocoder of HiFi-GAN for the synthesis of the audio waveform.

  \caption{Training and inference}
\end{algorithm}

\section{Experiments and Result}
% In this section, we will describe the experiment setup and the analysis of the results.
\subsection{Dataset}
\begin{figure}
    \centering
    
    \subfigure[Bathroom]{
    \includegraphics[width=0.45\linewidth]{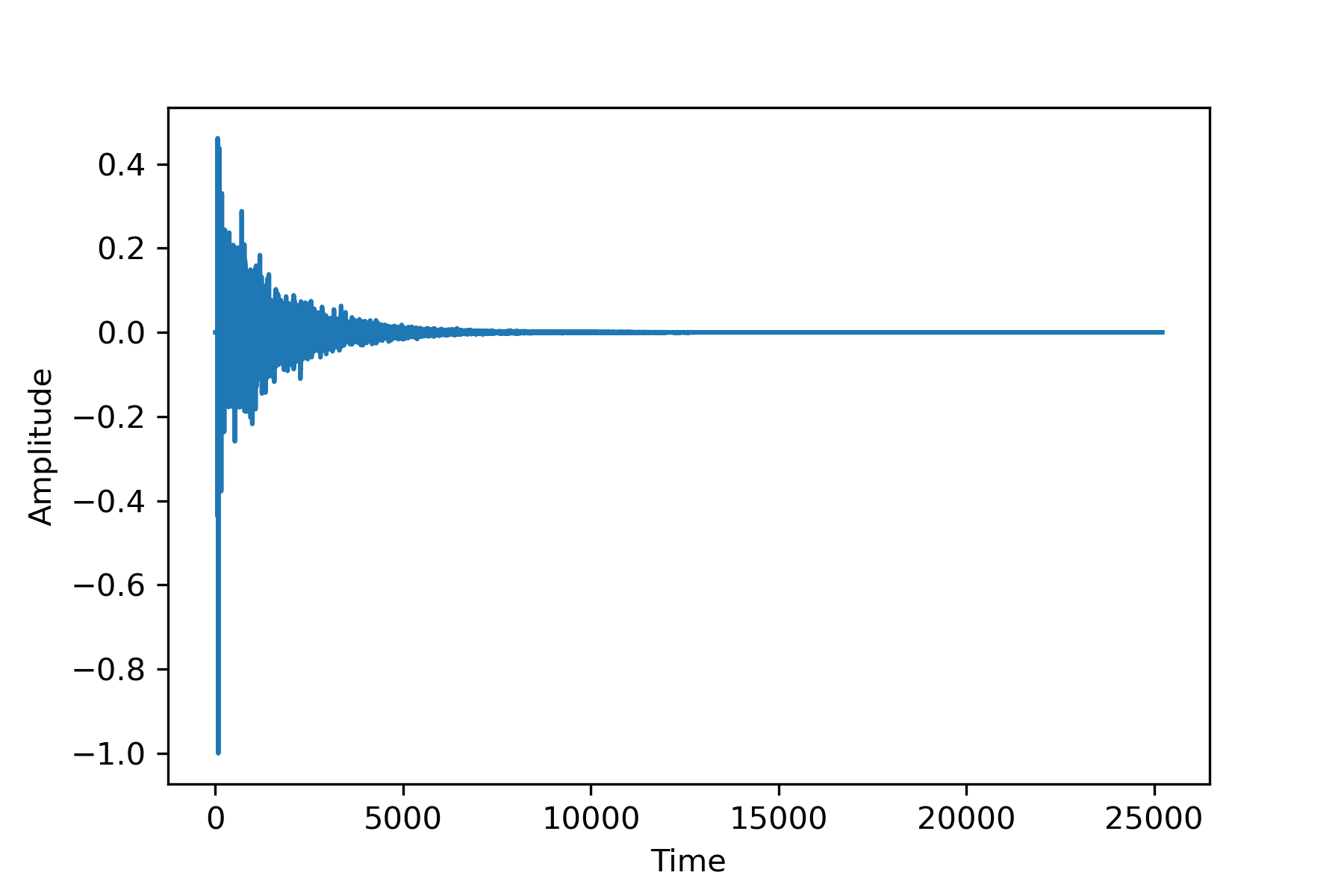}
    \label{result:rir Bathroom}
    }
    \subfigure[Cave]{
    \includegraphics[width=0.45\linewidth]{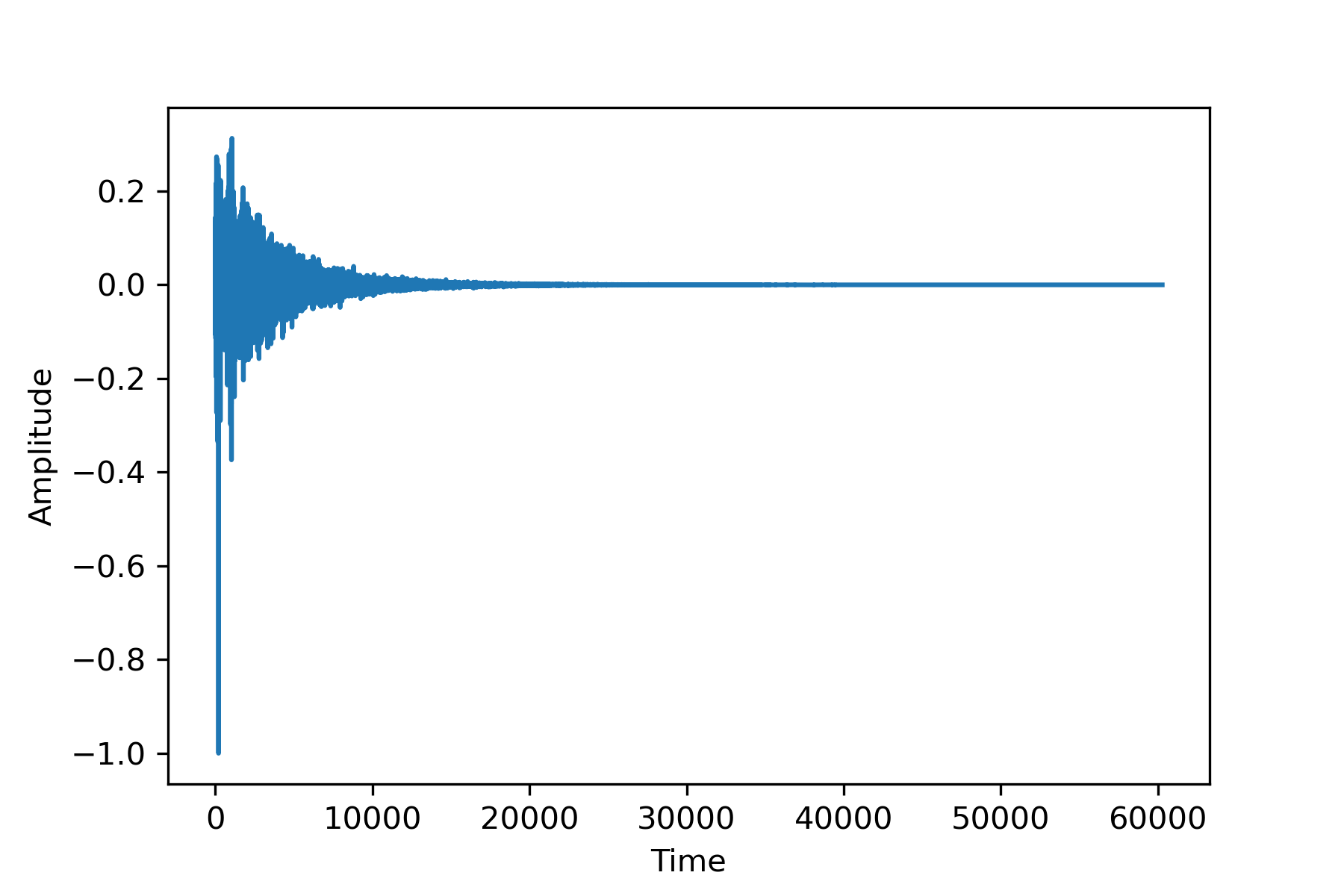}
    \label{result:rir Cave}
    }
    \subfigure[Classroom]{
    \includegraphics[width=0.45\linewidth]{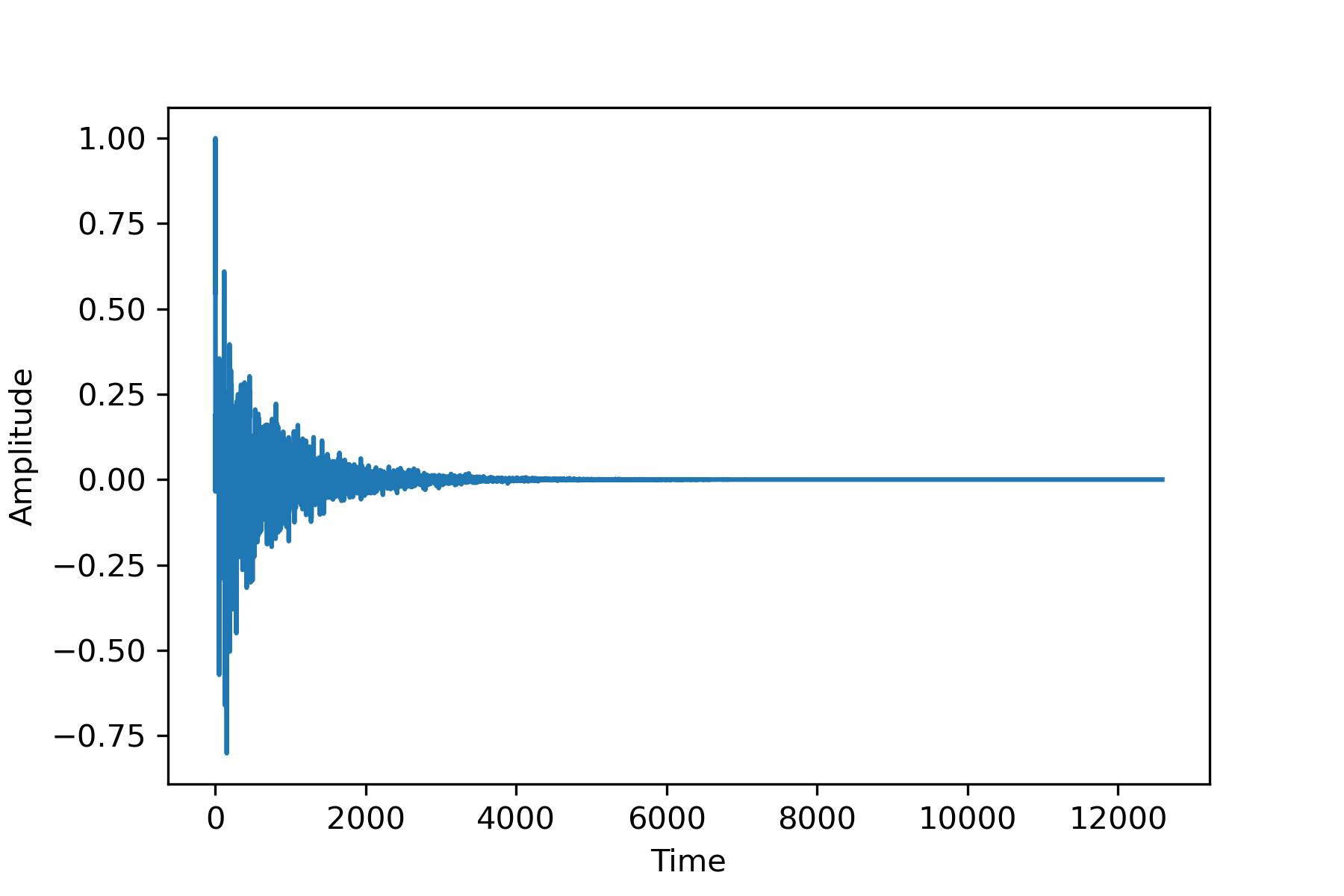}
    \label{result:rir Classroom}
    }
    \subfigure[Gallery]{
    \includegraphics[width=0.45\linewidth]{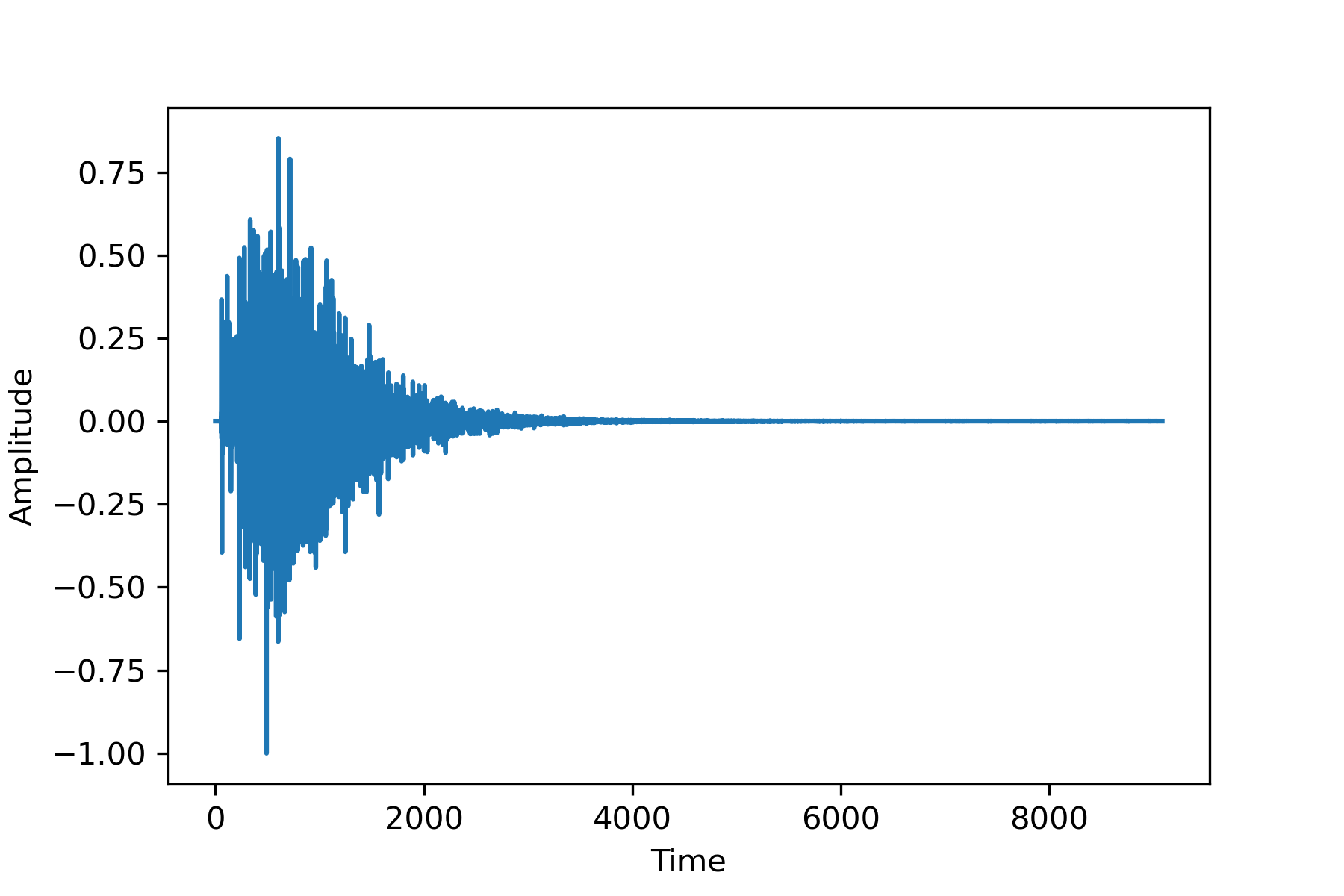}
    \label{result:rir Gallery}
    }
    \caption{The four room impulse response used for the environment simulation dataset: (a) Bathroom, (b) Cave, (c) Classroom and (d) Gallery}
    \label{fig:dataset rir}
\end{figure}

As the previous work by Ratnarajah \textit{et al.} \cite{ratnarajah2021ts}, they use a public dataset \cite{panayotov2015librispeech} in a far-field way to simulate the realistic data for training. We used four environment effects including \textit{Bathroom}, \textit{Cave}, \textit{Classroom} and \textit{Gallery} on the public dataset of LJSpeech~\cite{LJSpeech17} together to generate a dataset for the experiment. The room impulse response of the selected environment was convoluted with the raw speech in the LJSpeech. The four different environment room impulse responses were shown in Figure~\ref{fig:dataset rir}. The LJSpeech total has 13100 clips with a total duration of about 24 hours. We do a preprocess of that the environment effect was convolved with the audio wave to generate the simulated environment audio. After the preprocessing, we have four environment effects and finally enlarge the dataset five times of the LJSpeech.

\subsection{Experiment Setup}
As the method mainly do the conversion of environment effect from source speech to target, we compare the voice conversion models as baselines to do the task of environment effect switch. The baselines of AutoVC~\cite{qian2019autovc}, CycleGAN-VC3~\cite{kaneko2020cyclegan} and SpeechSplit~\cite{qian2020unsupervised} were retrained on the same dataset of LJSpeech for environment effect conversion.

The preprocess of mel spectrum, pitch, and energy was computed firstly. We used the pyworld for pitch estimation. All the audio data were resampled to 22.05kHz, the mel channels were set to 80. It should be noted that we chose the maximum length of the mel spectrum as 1200 for padding. 

As to the model configuration used in the experiment, we mainly based the backbone network of Fastspeech2~\cite{ren2020fastspeech}. We based on the main architecture of Fastspeech2 and alter the encoder for mel spectrum input with convolution 1D layer. The environment effect extractor was added with a U-net architecture convolution layer, which contains 4 down layers and 4 up layers. In the down layer, a 1D max pooling with kernel size of 2. In the up layer, a transposed 1D convolution layer, and a stack of 2 1D convolution layers same with the down layer. The gradient reversal layer was implemented within the backward function to do a negative process. The environment effect classifiers were both use a stack of two 1D convolutional layers, connect with a linear layer.

The training mainly contains a single GPU of Tesla V100. We set the batch size of 16, and total training steps of 900K, every 10K steps will save the trained model. We used Adam optimizer and set the $\beta_1$,$\beta_2$,$\epsilon$ to 0.9,0.98 and $10^{-9}$ respectively. 

\subsection{Objective Evaluations}

\begin{figure*}[htbp]
    \centering
    
    \subfigure[Bathroom]{
    \includegraphics[width=0.22\linewidth]{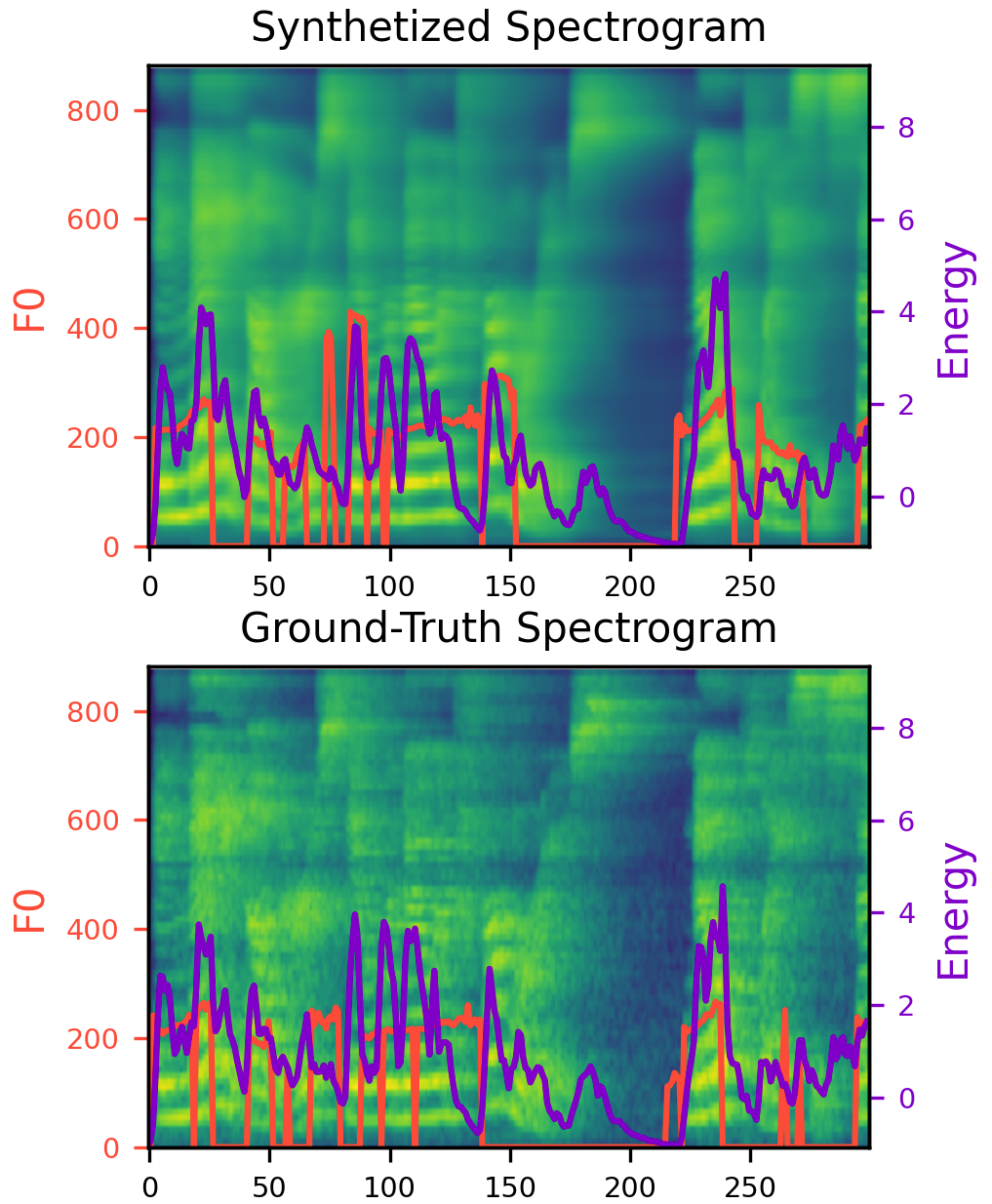}
    \label{result:spectrum Bathroom}
    }
    \subfigure[Cave]{
    \includegraphics[width=0.22\linewidth]{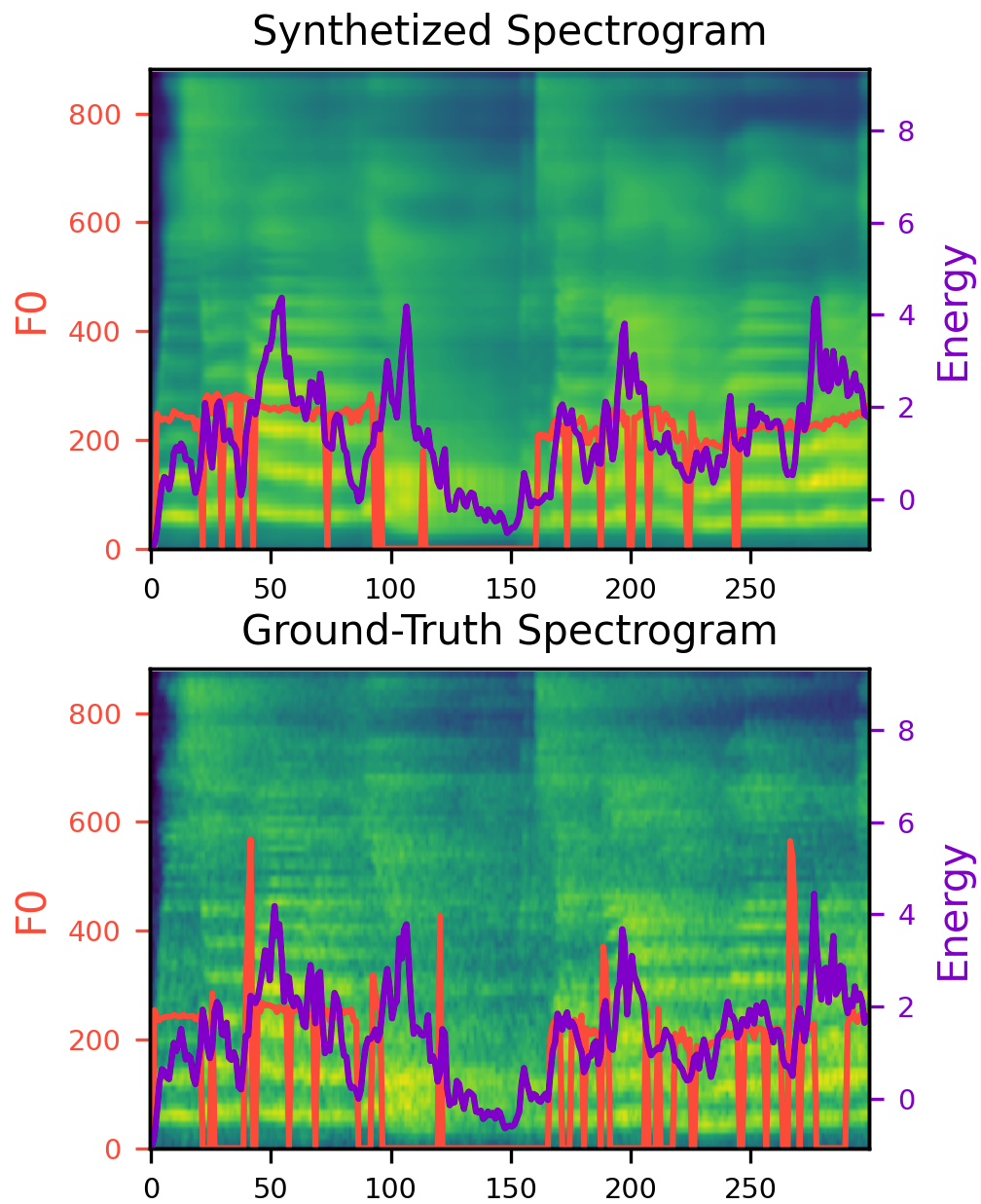}
    \label{result:spectrum Cave}
    }
    \subfigure[Classroom]{
    \includegraphics[width=0.22\linewidth]{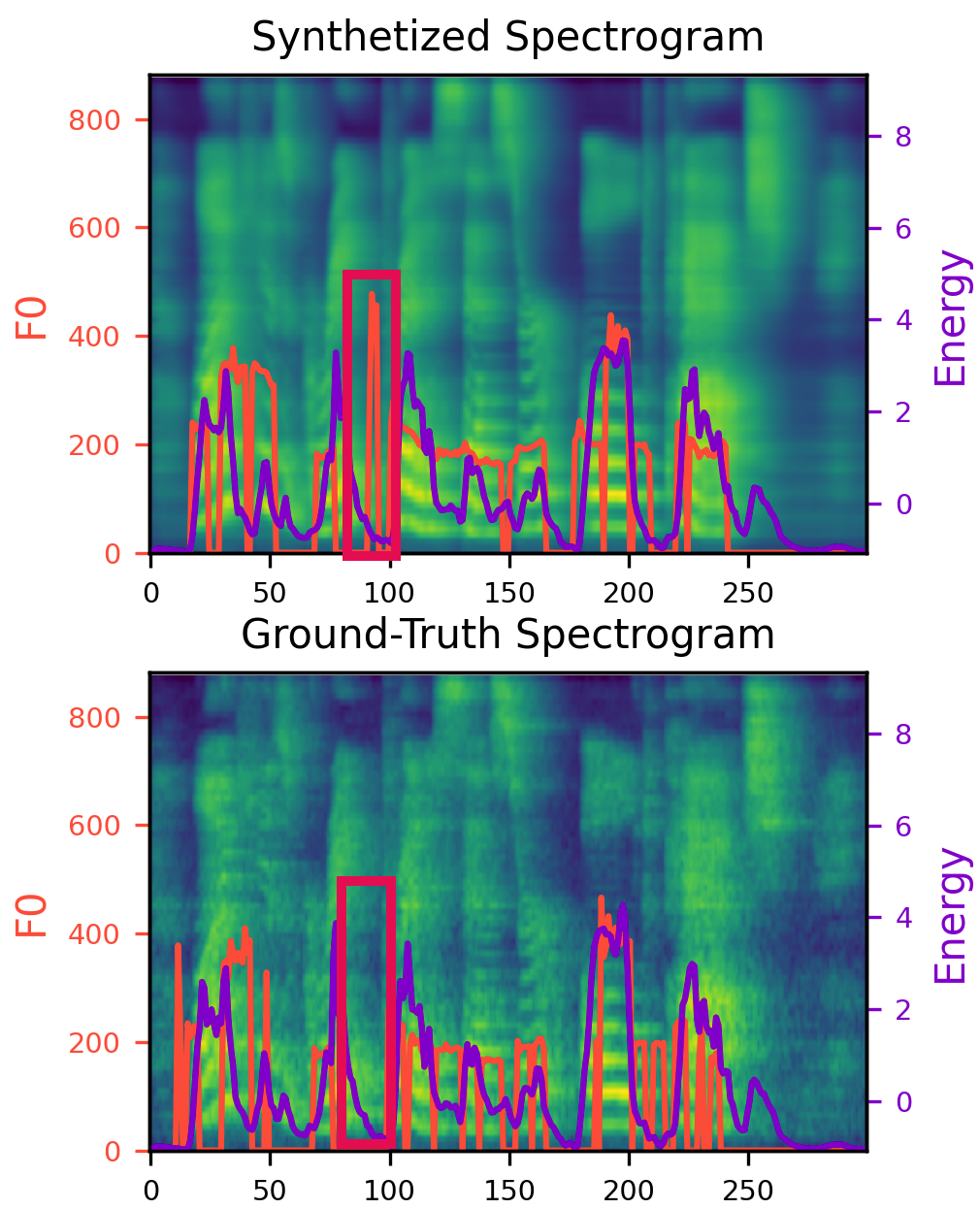}
    \label{result:spectrum Classroom}
    }
    \subfigure[Gallery]{
    \includegraphics[width=0.22\linewidth]{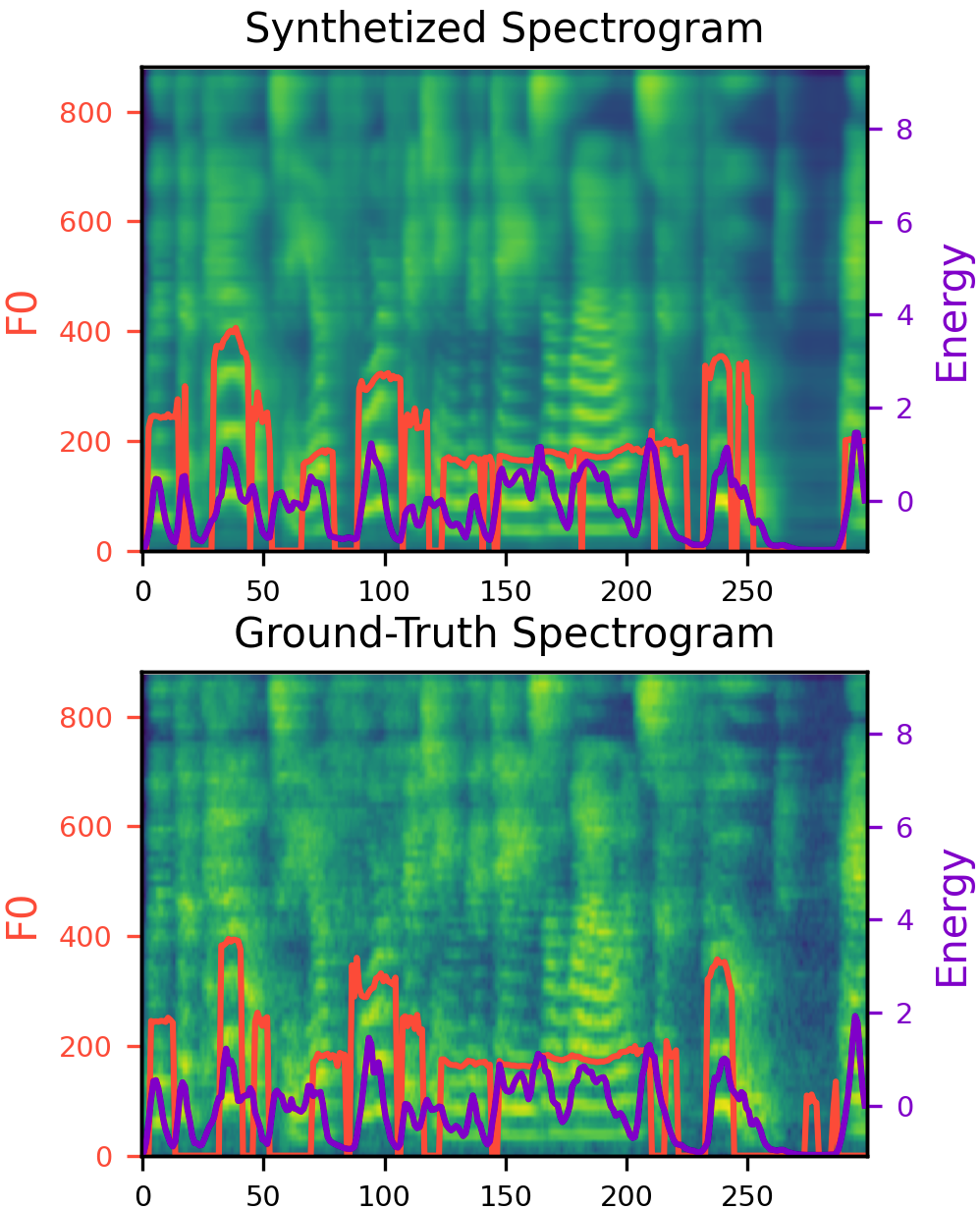}
    \label{result:spectrum Gallery}
    }
    \caption{The comparison of conversion speech with groudtruth in terms of spectrum, pitch and energy under different environment effect: (a) Bathroom, (b) Cave, (c) Classroom, (d) Gallery}
    \label{result:spectrum}
\end{figure*}

The Mel-cepstral distortion (MCD) was calculated between the generated speech with a specific environment and the groundtruth in Table~\ref{result:MCD tab}. The MCD was usually used in the voice conversion tasks, it measures the global structural difference. From the comparison results in terms of MCD, our model outperforms the baseline method under the environment of Bathroom, Cave, Classroom and Gallery. While the proposed method was slightly higher than the SpeechSplit for the environment conversion. The results may mean that all systems achieve comparable performance levels.

\begin{table}[htp]
	\caption{Comarision of MCD of different models. }
	\centering
	\label{result:MCD tab}
	  \begin{tabular}{ccccc}
	  \toprule
	  Method & Bathroom&Cave&Classroom&Gallery  \\
	  \midrule
	  AutoVC~\cite{qian2019autovc}  &  9.56 
	  &  9.32  
	  &  9.31 
	  & 9.29 \cr
	  CycleGAN-VC3~\cite{kaneko2020cyclegan} &  9.32 
	  & 9.02 
	   & 8.91 
	  & 9.14 \cr
	  SpeechSplit~\cite{qian2020unsupervised} & 8.62 
	  & 8.47 
	  &  8.40 
	  & 8.28 \cr
	  MetaSpeech &  \textbf{8.52 }
	  & \textbf{8.43 }
	  & \textbf{8.31 }
	  & \textbf{8.26} \cr
	  \bottomrule
	  \end{tabular}
  \end{table}

  \begin{table}[htp]
	\centering
	\caption{Comparison of MOS of different models.}
	\label{result:MOS tab} 
	  \begin{tabular}{ccccc}
	  \toprule
	  Method & Bathroom&Cave&Classroom&Gallery  \\
	  \midrule
	  Ground truth & 4.43 
	  & 4.52 
	  &4.60 & 4.58
	  \cr
	  AutoVC~\cite{qian2019autovc}  & 2.86 
	  & 3.08 
	  & 3.12 
	  & 3.20  \cr
	  CycleGAN-VC3~\cite{kaneko2020cyclegan} & 3.19 
	  & 3.29 
	  & 3.42 
	  & 3.24 \cr
	  SpeechSplit~\cite{qian2020unsupervised} & 3.58 
	  & 3.63 
	  & 3.76 
	  & 3.79 \cr
	  MetaSpeech & \textbf{3.60} 
	  & \textbf{3.68}
	  & \textbf{3.76} 
	  & \textbf{3.80}  \cr
	  \bottomrule
	  \end{tabular}
  \end{table}

 To more intuitively reflect the conversion of the different environmental effects. We further show conversion visualization of the spectrum, pitch, and energy under four different environments cases in Figure~\ref{result:spectrum}. 
 
The four pairs of the spectrum are randomly selected from the test set according to the different target environment effects. The pitch and the energy were plotted on the spectrum with the red line and purple line separately. From the comparison with the groundtruth, we can see the difference of mel spectrum, and the estimated pitch and energy. Focus on the environment of the Bathroom in Figure~\ref{result:spectrum Bathroom}, the synthesized spectrum is nearly the same as the groundtruth one. But there exists a blur region in high frequency band of the synthesized one. The other three environment target conversion has the same problem. It shows room for enhancement in the future. When we compare the pitch in the synthesized speech and the groundtruth, we can see there are little differences, such as the red rectangular box in Figure~\ref{result:spectrum Classroom}. The predicted energy of the synthesized speech and the groundtruth are nearly the same, and it can be proved from the Figure~\ref{result:spectrum Cave} and Figure~\ref{result:spectrum Gallery}. 

%  Since MCD and Mel spectrum cannot represent perception, we further conduct subjective evaluations on voice quality and environmental fitness. 

\subsection{Subjective Evaluations}

 In the subjective evaluation, we invited 10 listeners to evaluate the results.  For the MOS test, we randomly selected 5 speeches longer than 2 seconds and shorter than 6 seconds for each environment, and shuffle all the audio samples to let the tester give a score in the range 1-5 for each speech. We show them MOS test results in Table~\ref{result:MOS tab}. 

We selected three sentences for each environment speech, and each pair of audios contain conversion speech of the proposed method, the comparison methods audio alternate in the audio pairs. During the calculation of the result, we average the times of occurrence in the test. The listeners were asked which is more similar to the environment of the target speech \textit{X}. There are three options can be selected by the listeners, \textit{A}, \textit{B} and \textit{Fair}. We show the preference scores of environment effect in Table~\ref{ressult:abx}.

\begin{table}[htbp]
    \centering
    \caption{Comparison of preference score on similarity.}
    \label{ressult:abx}
    \begin{tabular}{ccccc}
    \toprule
         Method & Bathroom&Cave&Classroom&Gallery  \\
         \midrule
         AutoVC~\cite{qian2019autovc}& 0.08&0.06& 0.10&0.11\\
         CycleGAN-VC3~\cite{kaneko2020cyclegan}&0.22 &0.21&0.28 &0.23\\
         SpeechSplit~\cite{qian2020unsupervised}&0.32 &0.33&0.29 &0.31\\
         MetaSpeech& \textbf{0.38}&\textbf{0.40}&\textbf{0.33} &\textbf{0.35}\\
         \bottomrule
         
    \end{tabular}
    
\end{table}
The compared methods and the proposed all in a manner with no paired data to do the unsupervised reconstruction. From the result of the MOS test in Table~\ref{result:MOS tab}, we can see the proposed model could outperform the baseline methods under the environment of Bathroom, Cave, Classroom and Gallery. The proposed method outperforms all the baseline methods, while there is a big room with the groudtruth speech, the average score of our proposed method is about 3.7.  On one hand, it shows the validation of the proposed method and is comparable to the related works of voice conversion. Additionally, the conversion result under the different enviroment condition may lenarn the specifc environment reverberation. The baseline methods perform worse under the environment target, while the vocie conversion not with the goal of the environment effect conversion.

As shown in Table~\ref{ressult:abx}, the ABX test result on the similarity of target speech of specific environment effect revealed that the proposed method performed at a comparable level to the baseline method. There are 40\% choice will fall in the audio samples of the proposed method under the environment of Cave. While the similarity is slightly higher than baseline methods under the environment of Classroom. This result was consistent with the MCD objective evaluation and MOS evaluation.

\section{Conclusion}

% For more immerse to connect into Metaverse, voice conversion based method could convert the user speech to specific target speech with environment effect. While the voice conversion methods mainly focus on the conversion of speaker timbre.
In this paper, we proposed MetaSpeech, an environment effect conversion method containing an environment effect module to do a disentanglement of the environment effect with keeping the speaker timbre and the speech content. The speech conversion experiment is carried out by making a simulated environment effect dataset. The results show that the proposed method could do a valid conversion of the environment effect, and it outperforms the baseline methods from the voice conversion task in terms of MOS, MCD, and environment similarity. %The future work will pay attention to the control of the environment effect conversion without the reference environment speech.

\section{Acknowledgement}
This paper is supported by the Key Research and Development Program of Guangdong Province under grant No.2021B0101400003. Corresponding author is Jianzong Wang from Ping An Technology (Shenzhen) Co., Ltd (jzwang@188.com).

\bibliographystyle{IEEEtran}
\bibliography{mybib}

\begin{thebibliography}{10}
\providecommand{\url}[1]{#1}
\csname url@samestyle\endcsname
\providecommand{\newblock}{\relax}
\providecommand{\bibinfo}[2]{#2}
\providecommand{\BIBentrySTDinterwordspacing}{\spaceskip=0pt\relax}
\providecommand{\BIBentryALTinterwordstretchfactor}{4}
\providecommand{\BIBentryALTinterwordspacing}{\spaceskip=\fontdimen2\font plus
\BIBentryALTinterwordstretchfactor\fontdimen3\font minus
  \fontdimen4\font\relax}
\providecommand{\BIBforeignlanguage}[2]{{%
\expandafter\ifx\csname l@#1\endcsname\relax
\typeout{** WARNING: IEEEtran.bst: No hyphenation pattern has been}%
\typeout{** loaded for the language `#1'. Using the pattern for}%
\typeout{** the default language instead.}%
\else
\language=\csname l@#1\endcsname
\fi
#2}}
\providecommand{\BIBdecl}{\relax}
\BIBdecl

\bibitem{lee2021all}
L.-H. Lee, T.~Braud, P.~Zhou, L.~Wang, D.~Xu, Z.~Lin, A.~Kumar, C.~Bermejo, and
  P.~Hui, ``All one needs to know about metaverse: A complete survey on
  technological singularity, virtual ecosystem, and research agenda,''
  \emph{arXiv preprint arXiv:2110.05352}, 2021.

\bibitem{ning2021survey}
H.~Ning, H.~Wang, Y.~Lin, W.~Wang, S.~Dhelim, F.~Farha, J.~Ding, and
  M.~Daneshmand, ``A survey on metaverse: the state-of-the-art, technologies,
  applications, and challenges,'' \emph{arXiv preprint arXiv:2111.09673}, 2021.

\bibitem{lim2022realizing}
W.~Y.~B. Lim, Z.~Xiong, D.~Niyato, X.~Cao, C.~Miao, S.~Sun, and Q.~Yang,
  ``Realizing the metaverse with edge intelligence: A match made in heaven,''
  \emph{arXiv preprint arXiv:2201.01634}, 2022.

\bibitem{sparkes2021metaverse}
M.~Sparkes, ``What is a metaverse,'' 2021.

\bibitem{duan2021metaverse}
H.~Duan, J.~Li, S.~Fan, Z.~Lin, X.~Wu, and W.~Cai, ``Metaverse for social good:
  A university campus prototype,'' in \emph{Proceedings of the 29th ACM
  International Conference on Multimedia}, 2021, pp. 153--161.

\bibitem{ratnarajah2021fast}
A.~Ratnarajah, S.-X. Zhang, M.~Yu, Z.~Tang, D.~Manocha, and D.~Yu, ``Fast-rir:
  Fast neural diffuse room impulse response generator,'' \emph{arXiv preprint
  arXiv:2110.04057}, 2021.

\bibitem{SuwanaposeeGC22}
\BIBentryALTinterwordspacing
P.~Suwanaposee, C.~Gutwin, and A.~Cockburn, ``The influence of audio effects
  and attention on the perceived duration of interaction,'' \emph{Int. J. Hum.
  Comput. Stud.}, vol. 159, p. 102756, 2022. [Online]. Available:
  \url{https://doi.org/10.1016/j.ijhcs.2021.102756}
\BIBentrySTDinterwordspacing

\bibitem{RamirezWSB21}
\BIBentryALTinterwordspacing
M.~A.~M. Ram{\'{\i}}rez, O.~Wang, P.~Smaragdis, and N.~J. Bryan,
  ``Differentiable signal processing with black-box audio effects,'' in
  \emph{{IEEE} International Conference on Acoustics, Speech and Signal
  Processing, {ICASSP} 2021, Toronto, ON, Canada, June 6-11, 2021}.\hskip 1em
  plus 0.5em minus 0.4em\relax {IEEE}, 2021, pp. 66--70. [Online]. Available:
  \url{https://doi.org/10.1109/ICASSP39728.2021.9415103}
\BIBentrySTDinterwordspacing

\bibitem{martinez2021deep}
M.~A. Mart{\'\i}nez~Ram{\'\i}rez, ``Deep learning for audio effects modeling,''
  Ph.D. dissertation, Queen Mary University of London, 2021.

\bibitem{abs-2006-05584}
\BIBentryALTinterwordspacing
W.~Mitchell and S.~H. Hawley, ``Exploring quality and generalizability in
  parameterized neural audio effects,'' \emph{CoRR}, vol. abs/2006.05584, 2020.
  [Online]. Available: \url{https://arxiv.org/abs/2006.05584}
\BIBentrySTDinterwordspacing

\bibitem{bridges2020effects}
C.~R. Bridges~Jr, ``Effects of software tuning programs on vocal recordings,''
  in \emph{Audio Engineering Society Convention 149}.\hskip 1em plus 0.5em
  minus 0.4em\relax Audio Engineering Society, 2020.

\bibitem{canfield2018group}
E.~K. Canfield-Dafilou and J.~S. Abel, ``Group delay-based allpass filters for
  abstract sound synthesis and audio effects processing,'' in \emph{Proceedings
  of the 21st International Conference on Digital Audio Effects}, 2018.

\bibitem{gao2021vocal}
Y.~Gao, X.~Zhang, and W.~Li, ``Vocal melody extraction via hrnet-based singing
  voice separation and encoder-decoder-based f0 estimation,''
  \emph{Electronics}, vol.~10, no.~3, p. 298, 2021.

\bibitem{zhang2020research}
X.~Zhang, Y.~Yu, Y.~Gao, X.~Chen, and W.~Li, ``Research on singing voice
  detection based on a long-term recurrent convolutional network with vocal
  separation and temporal smoothing,'' \emph{Electronics}, vol.~9, no.~9, p.
  1458, 9 2020.

\bibitem{tang2022avqvc}
H.~Tang, X.~Zhang, J.~Wang, N.~Cheng, and J.~Xiao, ``Avqvc: One-shot voice
  conversion by vector quantization with applying contrastive learning,'' in
  \emph{ICASSP2022}.\hskip 1em plus 0.5em minus 0.4em\relax IEEE, 2022, pp.
  4613--4617.

\bibitem{ChenSH21}
\BIBentryALTinterwordspacing
M.~Chen, Y.~Shi, and T.~Hain, ``Towards low-resource stargan voice conversion
  using weight adaptive instance normalization,'' in \emph{{IEEE} International
  Conference on Acoustics, Speech and Signal Processing, {ICASSP} 2021,
  Toronto, ON, Canada, June 6-11, 2021}.\hskip 1em plus 0.5em minus 0.4em\relax
  {IEEE}, 2021, pp. 5949--5953. [Online]. Available:
  \url{https://doi.org/10.1109/ICASSP39728.2021.9415042}
\BIBentrySTDinterwordspacing

\bibitem{wang2022drvc}
Q.~Wang, X.~Zhang, J.~Wang, N.~Cheng, and J.~Xiao, ``Drvc: A framework of
  any-to-any voice conversion with self-supervised learning,'' in
  \emph{ICASSP2022}.\hskip 1em plus 0.5em minus 0.4em\relax IEEE, 2022, pp.
  3184--3188.

\bibitem{ChenWWL21}
Y.~Chen, D.~Wu, T.~Wu, and H.~Lee, ``Again-vc: {A} one-shot voice conversion
  using activation guidance and adaptive instance normalization,'' in
  \emph{{ICASSP} 2021}.\hskip 1em plus 0.5em minus 0.4em\relax {IEEE}, 2021,
  pp. 5954--5958.

\bibitem{asru2021tang}
H.~Tang, X.~Zhang, J.~Wang, N.~Cheng, Z.~Zeng, E.~Xiao, and J.~Xiao, ``{TGAVC}:
  Improving autoencoder voice conversion with text-guided and adversarial
  training,'' in \emph{ASRU2021}.\hskip 1em plus 0.5em minus 0.4em\relax
  {IEEE}, 2021, pp. 1--6.

\bibitem{HayashiHKT21}
T.~Hayashi, W.~Huang, K.~Kobayashi, and T.~Toda, ``Non-autoregressive
  sequence-to-sequence voice conversion,'' in \emph{{IEEE} International
  Conference on Acoustics, Speech and Signal Processing, {ICASSP} 2021,
  Toronto, ON, Canada, June 6-11, 2021}.\hskip 1em plus 0.5em minus 0.4em\relax
  {IEEE}, 2021, pp. 7068--7072.

\bibitem{asru2021zhang}
X.~Zhang, J.~Wang, N.~Cheng, E.~Xiao, and J.~Xiao, ``{CycleGEAN}:cycle
  generative enhanced adversarial network for voice conversion,'' in
  \emph{ASRU2021}.\hskip 1em plus 0.5em minus 0.4em\relax {IEEE}, 2021, pp.
  1--6.

\bibitem{kaneko2017parallel}
T.~Kaneko and H.~Kameoka, ``Parallel-data-free voice conversion using
  cycle-consistent adversarial networks,'' \emph{arXiv preprint
  arXiv:1711.11293}, 2017.

\bibitem{kameoka2018stargan}
H.~Kameoka, T.~Kaneko, K.~Tanaka, and N.~Hojo, ``Stargan-vc: Non-parallel
  many-to-many voice conversion using star generative adversarial networks,''
  in \emph{2018 IEEE Spoken Language Technology Workshop (SLT)}.\hskip 1em plus
  0.5em minus 0.4em\relax IEEE, 2018, pp. 266--273.

\bibitem{kaneko2021maskcyclegan}
T.~Kaneko, H.~Kameoka, K.~Tanaka, and N.~Hojo, ``Maskcyclegan-vc: Learning
  non-parallel voice conversion with filling in frames,'' in \emph{ICASSP
  2021-2021 IEEE International Conference on Acoustics, Speech and Signal
  Processing (ICASSP)}.\hskip 1em plus 0.5em minus 0.4em\relax IEEE, 2021, pp.
  5919--5923.

\bibitem{zhang2021singer}
X.~Zhang, J.~Qian, Y.~Yu, Y.~Sun, and W.~Li, ``Singer identification using deep
  timbre feature learning with knn-net,'' in \emph{ICASSP2021}.\hskip 1em plus
  0.5em minus 0.4em\relax IEEE, 2021, pp. 3380--3384.

\bibitem{hsu2016voice}
C.-C. Hsu, H.-T. Hwang, Y.-C. Wu, Y.~Tsao, and H.-M. Wang, ``Voice conversion
  from non-parallel corpora using variational auto-encoder,'' in \emph{2016
  Asia-Pacific Signal and Information Processing Association Annual Summit and
  Conference (APSIPA)}.\hskip 1em plus 0.5em minus 0.4em\relax IEEE, 2016, pp.
  1--6.

\bibitem{kameoka2018acvae}
H.~Kameoka, T.~Kaneko, K.~Tanaka, and N.~Hojo, ``Acvae-vc: Non-parallel
  many-to-many voice conversion with auxiliary classifier variational
  autoencoder,'' \emph{arXiv preprint arXiv:1808.05092}, 2018.

\bibitem{zhang2020singing}
X.~Zhang, S.~Li, Z.~Li, S.~Chen, Y.~Gao, and W.~Li, ``Singing voice detection
  using multi-feature deep fusion with cnn,'' in \emph{CSMT2019}.\hskip 1em
  plus 0.5em minus 0.4em\relax Springer, 2020, pp. 41--52.

\bibitem{qian2019autovc}
K.~Qian, Y.~Zhang, S.~Chang, X.~Yang, and M.~Hasegawa-Johnson, ``Autovc:
  Zero-shot voice style transfer with only autoencoder loss,'' in
  \emph{International Conference on Machine Learning}.\hskip 1em plus 0.5em
  minus 0.4em\relax PMLR, 2019, pp. 5210--5219.

\bibitem{qian2020unsupervised}
K.~Qian, Y.~Zhang, S.~Chang, M.~Hasegawa-Johnson, and D.~Cox, ``Unsupervised
  speech decomposition via triple information bottleneck,'' in
  \emph{International Conference on Machine Learning}.\hskip 1em plus 0.5em
  minus 0.4em\relax PMLR, 2020, pp. 7836--7846.

\bibitem{qian2021global}
K.~Qian, Y.~Zhang, S.~Chang, J.~Xiong, C.~Gan, D.~Cox, and M.~Hasegawa-Johnson,
  ``Global rhythm style transfer without text transcriptions,'' \emph{arXiv
  preprint arXiv:2106.08519}, 2021.

\bibitem{kaneko2019cyclegan}
T.~Kaneko, H.~Kameoka, K.~Tanaka, and N.~Hojo, ``Cyclegan-vc2: Improved
  cyclegan-based non-parallel voice conversion,'' in \emph{ICASSP 2019-2019
  IEEE International Conference on Acoustics, Speech and Signal Processing
  (ICASSP)}.\hskip 1em plus 0.5em minus 0.4em\relax IEEE, 2019, pp. 6820--6824.

\bibitem{zhao2022nnspeech}
B.~Zhao, X.~Zhang, J.~Wang, N.~Cheng, and J.~Xiao, ``nnspeech: Speaker-guided
  conditional variational autoencoder for zero-shot multi-speaker
  text-to-speech,'' in \emph{ICASSP2022}.\hskip 1em plus 0.5em minus
  0.4em\relax IEEE, 2022, pp. 4293--4297.

\bibitem{ren2020fastspeech}
Y.~Ren, C.~Hu, X.~Tan, T.~Qin, S.~Zhao, Z.~Zhao, and T.-Y. Liu, ``Fastspeech 2:
  Fast and high-quality end-to-end text to speech,'' \emph{arXiv preprint
  arXiv:2006.04558}, 2020.

\bibitem{ratnarajah2021ts}
A.~Ratnarajah, Z.~Tang, and D.~Manocha, ``Ts-rir: Translated synthetic room
  impulse responses for speech augmentation,'' \emph{arXiv preprint
  arXiv:2103.16804}, 2021.

\bibitem{panayotov2015librispeech}
V.~Panayotov, G.~Chen, D.~Povey, and S.~Khudanpur, ``Librispeech: an asr corpus
  based on public domain audio books,'' in \emph{2015 IEEE international
  conference on acoustics, speech and signal processing (ICASSP)}.\hskip 1em
  plus 0.5em minus 0.4em\relax IEEE, 2015, pp. 5206--5210.

\bibitem{LJSpeech17}
K.~Ito and L.~Johnson, ``The lj speech dataset,''
  \url{https://keithito.com/LJ-Speech-Dataset/}, 2017.

\bibitem{kaneko2020cyclegan}
T.~Kaneko and et~al., ``Cyclegan-vc3: Examining and improving cyclegan-vcs for
  mel-spectrogram conversion,'' \emph{arXiv preprint arXiv:2010.11672}, 2020.

\end{thebibliography}
\end{document}